\documentclass[onecolumn,usenatbib,letter]{mn2e}
\usepackage{epsfig,amsmath,amssymb,amsfonts,mathrsfs,latexsym,graphicx,deluxetable}
\begin{document}

\author[Watts \& Strohmayer]{Anna L. Watts$^1$ and Tod E. Strohmayer$^2$
\\ $^1$ Max Planck Institut f\"ur
  Astrophysik, Karl-Schwarzschild-Str.1, 85741 Garching, Germany; anna@mpa-garching.mpg.de \\
$^2$ Astrophysics Science Division,
NASA Goddard Space Flight Center, Greenbelt, MD 20771, USA; stroh@milkyway.gsfc.nasa.gov}

\title[XTE J1814-338 burst oscillation energy dependence]{The energy dependence of burst oscillations from the accreting
millisecond pulsar XTE J1814-338}

\maketitle

\begin{abstract}

The nature of the asymmetry that gives rise to Type I X-ray burst
oscillations on accreting neutron stars remains a matter of
debate.  Of particular interest is whether the burst oscillation
mechanism differs between the bursting millisecond pulsars and the
non-pulsing systems. One means to diagnose this is to study the energy
dependence of the burst oscillations: here we present an analysis of
 oscillations from 28 bursts observed during the 2003 outburst of the
accreting millisecond pulsar XTE J1814-338.  We find that the 
fractional amplitude of the burst oscillations falls with energy, in
contrast to the behaviour found by Muno et al. (2003) in the burst
oscillations from a set of non-pulsing systems.  The drop with energy mirrors that seen in the 
accretion-powered pulsations; in this respect XTE J1814-338 behaves
like the other accreting millisecond pulsars.  The burst
oscillations show no evidence for either hard or soft lags, in
contrast to the persistent pulsations, which show soft lags of up to
50 $\mu$s. The fall in amplitude with energy is inconsistent with
current surface mode and simple hot spot models of 
burst oscillations.   We discuss improvements to the models and
uncertainties in the physics that might resolve these issues.  
\end{abstract}


\section{Introduction}
\label{intro}

The neutron star Low Mass X-ray Binaries (LMXBs) are neutron stars accreting from low-mass companions that over-fill their Roche lobes.  Of these systems some, the accreting millisecond pulsars, are also observed as X-ray pulsars.  
Of the $\approx$ 160 known non-pulsing LMXBs, nearly half exhibit Type I X-ray bursts due to
unstable nuclear 
burning of accreting material.  High frequency modulations of the
X-ray lightcurve, known as burst oscillations, have now been detected in the
bursts from twelve of these systems (see the recent review by \citet{str06}).
Of the seven known accreting millisecond pulsars, 
three exhibit bursts.  Two of these systems, SAX J1808.4-3658
(hereafter J1808) and XTE J1814-338 (hereafter J1814), have burst
oscillations. 

For the pulsars, the burst oscillation frequency is at
or very close to the known spin frequency of the star \citep{cha03,
str03}.  This suggests 
that the rotation modulates an asymmetry on the burning surface that
is near-stationary in the corotating frame.  For the non-pulsars the
situation is less clear, because there is no independent measure of
the spin frequency.  The fact that the frequency is highly stable from
burst to burst,
however \citep{str98, mun02}, implies that there is at least a strong
dependence on stellar spin.    The
detection of highly coherent oscillations lasting several hundred
seconds during a superburst from the LMXB 4U 1636-356 adds further
support to this hypothesis \citep{str02a}.  

The precise nature of the brightness asymmetry is as yet unclear.  
The simplest possibility is uneven distribution of fuel, leading to
patchy burning.  For the pulsars we know 
that initial fuel distribution is 
asymmetric, but the fuel could spread rapidly
over the surface.  For
the non-pulsars there is no evidence of asymmetric fuel deposition,
although there are several mechanisms that might render this process
(and the associated pulsations)
undetectable 
 \citep{bra87, kyl87, woo88, mes88, cum01, tit02}. Alternate
mechanisms that do not rely on asymmetric fuel distribution 
include the development of  
vortices driven by the Coriolis force \citep{spi02},  or global modes
in the burning 
surface layers \citep{mcd87, lee04,
  hey04, lee05, cum05, pir05a}.  

One means of diagnosing the nature of the asymmetry is to study the
energy dependence of the burst oscillations.  One can study both the dependence of amplitude on energy, and the relative phases of the waveforms as a function of energy.  A ``soft lag'' implies that the soft (low energy) pulse arrives later in phase than the hard (high energy) pulse, and vice versa.  \citet{mun03} examined  
the energy dependence expected for a simple hot spot model, generating simulated
 {\it Rossi X-ray Timing Explorer} (RXTE) Proportional Counter
Array (PCA) lightcurves for neutron stars with a circular uniform
temperature hot spot on a cooler background. The temperature contrast between the spot and the rest of the star, coupled with Doppler effects due to the star's rotation, leads naturally to a rise in fractional amplitude with 
energy in the 3-20 keV band.  The hot spot model also predicts soft
lags.  Models where the burst oscillations are caused by surface modes
 also predict a rise in fractional amplitude
with energy \citep{hey05, pir06}.  In most cases the mode models
predict soft lags, although \citet{lee05} found hard lags for a
restricted subset of 
possible geometries.  

The most comprehensive
study to date of burst oscillations from the non-pulsing LMXBs, using
RXTE PCA data, found that fractional
amplitude rises with energy over the 3-20 keV energy band \citep{mun03}.
The authors show that the observations
are compatible with the simple hot spot model if the bulk of the
stellar surface (away from the hot spot) emits in the lower part of the
PCA energy band.  The data are also compatible with surface mode models
\citep{pir06}, provided that the required amplitudes can be excited.

Muno et al. also examined whether there were
phase lags between soft and hard photons.  For most bursts the data
were consistent with there being no phase lag; for 13 of 54 bursts
studied the
data were inconsistent at the 90\% confidence level with being
constant.  In these bursts the trend was for the hard photons to lag
the soft photons by up to 0.12 cycles, whereas both simple hot spot and mode models
predict soft lags.  The authors considered several
mechanisms that might reverse a soft lag.  They concluded that
inverse Compton scattering of soft photons to higher energies by a hot
corona of electrons was the most likely 
candidate \citep{mil95}.  The observational implications of such a
corona, however, and its behaviour during a burst, remain to be worked
out in detail.

To date there has been no study reporting the energy dependence of the burst
oscillations of the accreting millisecond
pulsars.  Given the
possibility that the burst oscillation mechanism may differ from that
operating in the non-pulsing sources, this is clearly an omission.  In
this paper we attempt to rectify this by presenting an analysis of the
energy dependence of the burst oscillations of J1814.  
 
The source  was
first detected in outburst on 2003 June 3 by the RXTE Galactic bulge
monitoring campaign.  Its status as a millisecond pulsar was confirmed
by a longer observation on June 5 \citep{mar03}. The pulsar has a spin
frequency of 314.36 Hz, resides in a binary with orbital period 
4.275 hr, and has a minimum companion mass of $\approx 0.15 M_\odot$
\citep{mar03b}.  The binary orbit is the widest of the seven known accreting
millisecond pulsars.
J1814 remained in outburst 
until mid-July 2003, and in this time 28 Type I X-ray bursts were
observed, all with detectable burst oscillations at the spin
frequency \citep{str03}.  The bursts show significant harmonic
content, which could be used to constrain the equation of state and
the system geometry
\citep{bha05}.  A detailed study by \citet{wat05} found that there is no evidence for fractional amplitude
variation or frequency shifts in any of the bursts apart from the one
burst 
that appears to show photospheric radius expansion.  For
most of the bursts, fractional amplitude is 
consistent with that of the persistent pulsations,
although there is a small population of bursts with amplitudes that are
substantially lower.   For the first harmonic,
substantial differences between the burst and accretion-powered
oscillations indicate that hot spot geometry is not the only factor
giving rise to harmonic content in the latter.  There are however no
detectable phase 
shifts between the burst and accretion-powered pulsations
\citep{str03}, suggesting a model where the presence of the magnetic field somehow leads to a temperature asymmetry centered on the polar cap when the burst ignites (by allowing additional fuel build-up, for example).  In this
complementary study we test the models further by examining the
energy dependence of the burst oscillations, comparing the
behaviour to that of the accretion-powered pulsations.  Section \ref{data} details our method of
analysis and our results.  In Section \ref{disc}
we discuss our findings in the light of current burst oscillation
models, and outline areas for future theoretical study.  

\section{Data analysis}
\label{data}

\subsection{Methodology}
\label{method}

Almost all of the PCA data for the outburst is event mode data with  125
$\mu$s time resolution and 64 energy bins covering the range 2 -
120 keV.  The exception is the first burst, which was recorded in
GoodXenon mode, which has higher time and energy resolution (1$\mu$s
and 256 energy bins). Event mode data overruns, which are often seen in the
bursts of brighter sources, were not seen in any of the J1814 bursts.
The data were barycentered prior to analysis, using the JPL DE405 ephemeris and the
source position determined from PCA scans \citep{mar03, kra05}.    

Computation of fractional amplitudes is done in two ways.  We start
by using the $Z^2_n$ statistic \citep{buc83, str02b}.
This measure is very similar to the standard power spectrum
computed from a Fourier transform, but does not require that the event
data be binned.  It is defined as

\begin{equation}
Z^2_n = \frac{2}{N} \sum_{k=1}^n\left[\left(\sum_{j=1}^N \cos
k\phi_j\right)^2 + \left(\sum_{j=1}^N \sin
k\phi_j\right)^2 \right]
\end{equation}
where $n$ is the number of harmonics summed (we use $n=1$ throughout
this paper), $N$ is the total
number of photons, and $j$ is an index applied to each photon.  The
phase $\phi_j$ calculated for each photon is

\begin{equation}
\phi_j = 2\pi \int_{t_0}^{t_1} \nu(t) dt
\end{equation}
where $\nu(t)$ is the frequency model and $t_j$ is the arrival time of the photon
relative to some reference time.  We use a frequency model in which
the intrinsic spin rate is modified by orbital Doppler
shifts using the best fit orbital ephemeris.  Given a measured
$Z^2_n$, which we 
call $Z_m$, the probability of the true signal power lying between 0 and $Z_s$
 is given by:

\begin{equation}
f(Z_s: Z_m) = \exp\left[ -\frac{Z_m + Z_s}{2}\right]
\left[\sum_{k=0}^\infty \sum_{l=0}^{k + n - 1} \frac{(Z_s)^k
(Z_m)^l}{l!k!2^{k+l}}\right]
\end{equation}
\citep{gro75, vau94}. We take the best estimate for $Z_s$ to be that
for which $f(Z_s: Z_m) = 0.5$.  Given this $Z_s$ the RMS fractional
amplitude $r$ is then given by 

\begin{equation}
r = \left(\frac{Z_s}{N}\right)^{1/2} \left(\frac{N}{N - N_b}\right)
\end{equation}
where $N_b$ is the number of photons due to background accumulated in
the energy band of interest during the observation period.  $N_b$ is
estimated using the standard FTOOLS routine pcabackest and the PCA
background models.  

The second method used is to generate a folded pulse profile (using
the frequency model) and then fit a sinusoidal model with as many
harmonics as necessary.  We use this
method to check the fractional amplitudes computed using the $Z^2_n$
statistic.  Pulse profile fitting also allows us to calculate phase shifts between the
different energy bands, something that is not possible using the
$Z^2_n$ statistic alone.  A more extensive discussion of both of these
methods is given in Section 2.1 of \citet{wat05}.  

The other issue to consider when computing fractional amplitudes for
pulsar burst oscillations is that the accretion process may continue
during the bursts.  If this is the case then the measured fractional
amplitude will contain contributions from both the burst process and
the accretion process:

\begin{equation}
\label{prop}
r = \frac{r_\mathrm{bur} N_\mathrm{bur} + r_\mathrm{acc}
N_\mathrm{acc}}{N_s}
\end{equation}
$N_\mathrm{bur}$ and $N_\mathrm{acc}$ are the number of source photons
arising from the burst and accretion processes respectively, with
$r_\mathrm{bur}$ and $r_\mathrm{acc}$ being the fractional amplitudes
of the two different processes.  The total number of source photons is
$N_s = N_\mathrm{bur} + N_\mathrm{acc}$.  Since we are not always in
the regime where $N_\mathrm{bur} \gg N_\mathrm{acc}$ we will need to
estimate $N_\mathrm{acc}$ and $r_\mathrm{acc}$ in order to check
whether $r_\mathrm{bur}$ differs substantially from that measured.  We
do this in Section \ref{acc}.  

\subsection{Accretion-powered pulsations}
\label{acc}

\begin{figure}
\centering
\includegraphics[clip]{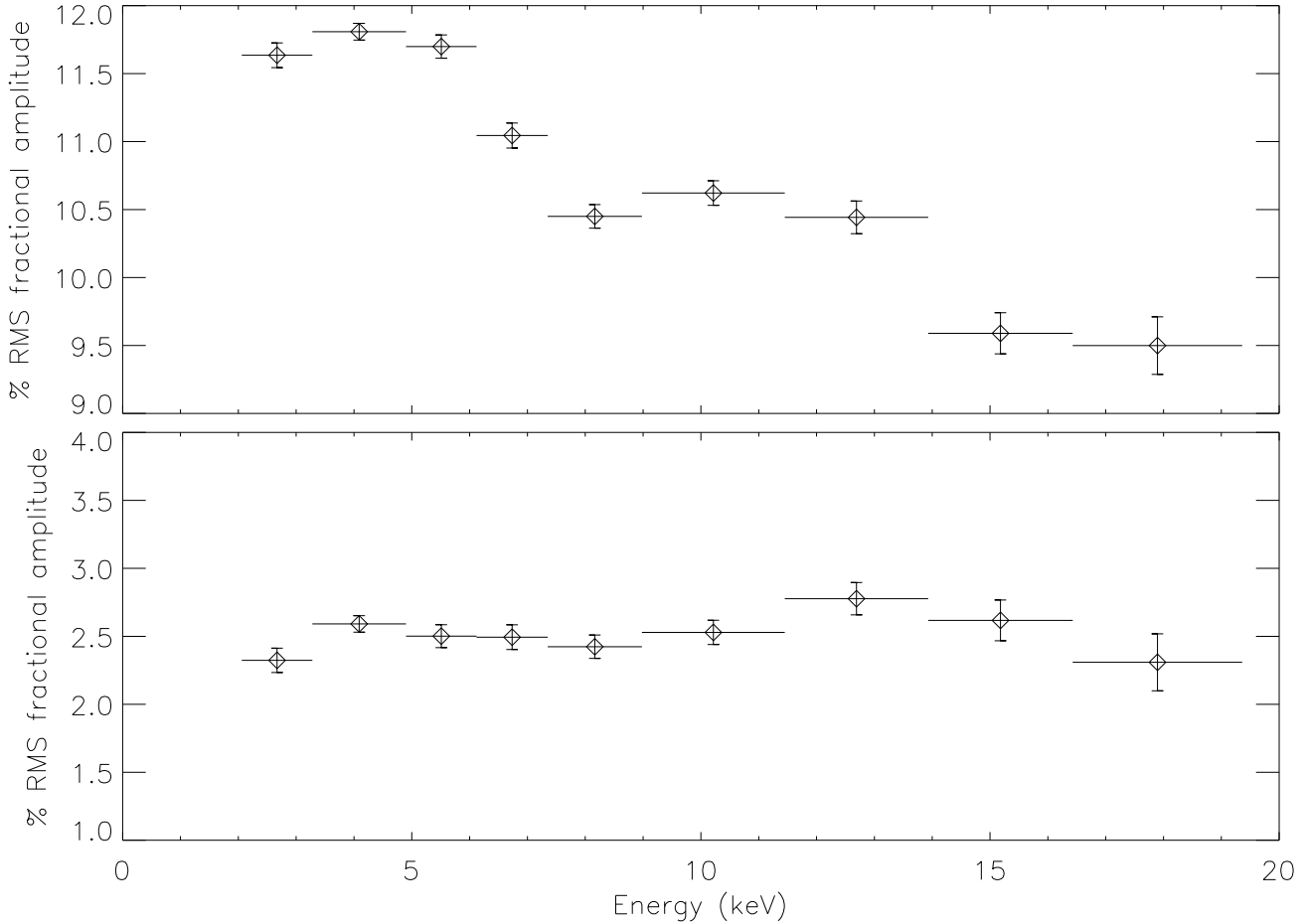}
\caption{Dependence of fractional amplitude on energy for the
accretion-powered pulsations, computed from a folded profile using
data from the whole outburst.  The fundamental (top panel) shows a
clear drop in amplitude with energy.  The amplitude of the first harmonic, by contrast (lower panel) shows little change with energy.  }  
\label{fa}
\end{figure}

\begin{figure}
\begin{center}
\includegraphics[clip]{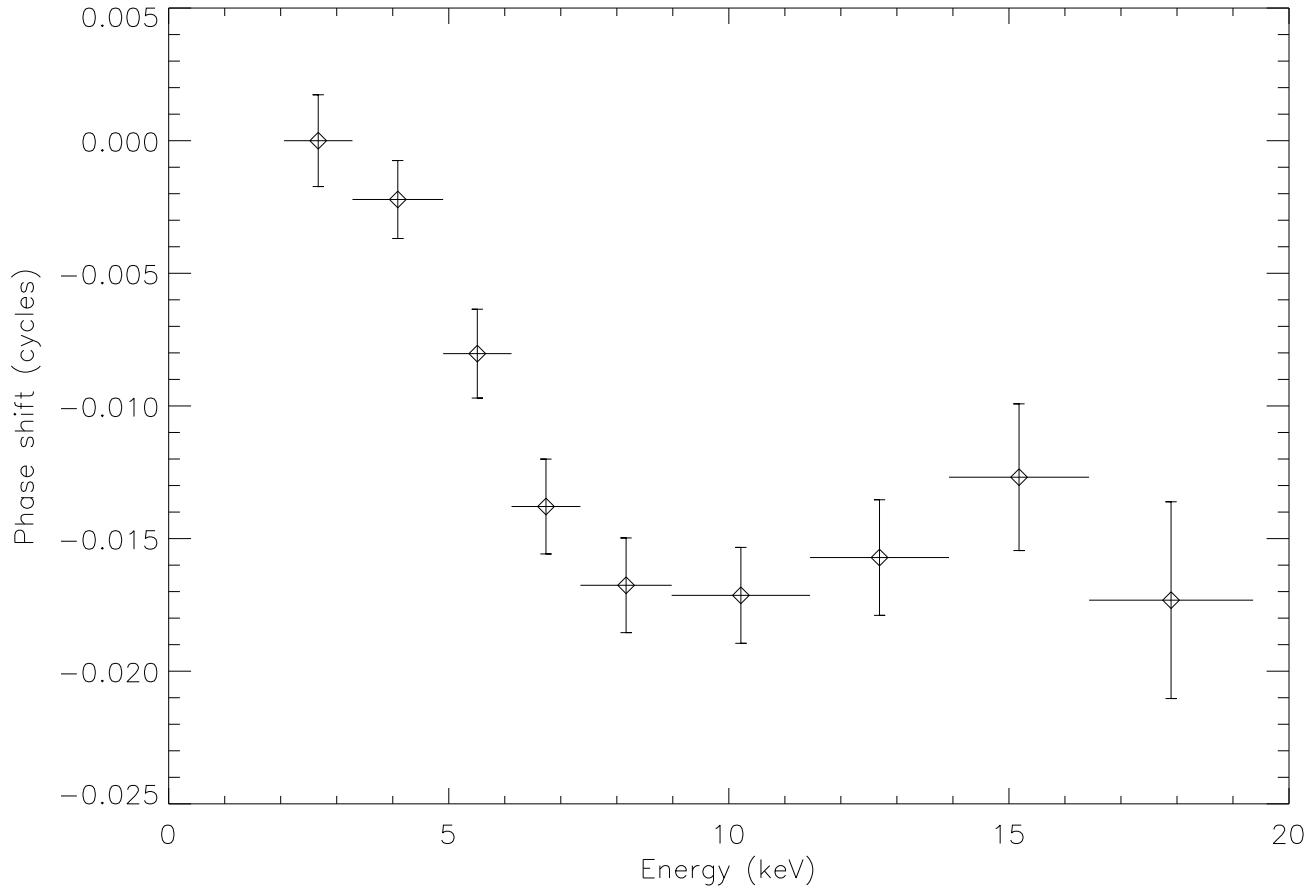}
\end{center}
\caption{Phase shifts compared to the phase for the lowest energy band
(2-3.3 keV) for
the accretion-powered pulsations.
Negative values indicate that the pulse profile in the energy band being
studied leads the profile in the lowest energy band, with
0.01 cycles corresponding to 31.8$\mu$s.  The phase shifts were
computed by
folding together data from the whole outburst.}  
\label{fb}
\end{figure}

We start by folding together data from the
whole outburst to generate a high resolution plot of amplitude and
phase lags against energy for the accretion-powered pulsations.  Figure \ref{fa} shows the behaviour of
amplitude against energy, for both the fundamental and first
harmonic.  There is a clear drop in amplitude with energy for the
fundamental, of about 2\% RMS over the 2-20 keV band.  The
amplitude of the first harmonic, by contrast, varies little with
energy.  Figure \ref{fb} shows the phase shifts
between the different energy bands for the fundamental.  Soft lags of
up to 50 $\mu$s ($\approx$ 0.015 cycles) develop over the range 2-7 keV, with the lags levelling
off at higher energies.  The same behaviour is seen in the phase shifts
computed using the first harmonic.  

This behaviour mirrors that
seen in the accretion-powered 
pulsations of the other millisecond pulsars:  J1808 \citep{cui98,
gie02}, XTE J0929-314 \citep{gal02}, XTE J1807-294 
\citep{kir04},  IGR
J00291+5934 \citep{gal05} and XTE J1751-305 \citep{gie05} all show
 a drop of 
fractional amplitude with energy within the PCA energy band. The other
sources also show soft lags,  although at only  
50 $\mu$s the J1814 lags are smaller than those measured for 
 the other systems.  Detailed spectral modelling for J1808 by
\citet{gil98}, \citet{gie02}  
and \citet{pou03} suggests that the amplitude drop and lags can be explained by the presence of a hard Comptonized component due to
boundary layer emission together with a softer blackbody component due to a
hot spot at the footpoint of the accretion channel.  Similar detailed modelling for J1814, which has a harder spectrum than 
J1808 \citep{mar03b}, has yet to be done.

\begin{figure}
\begin{center}
\includegraphics[clip]{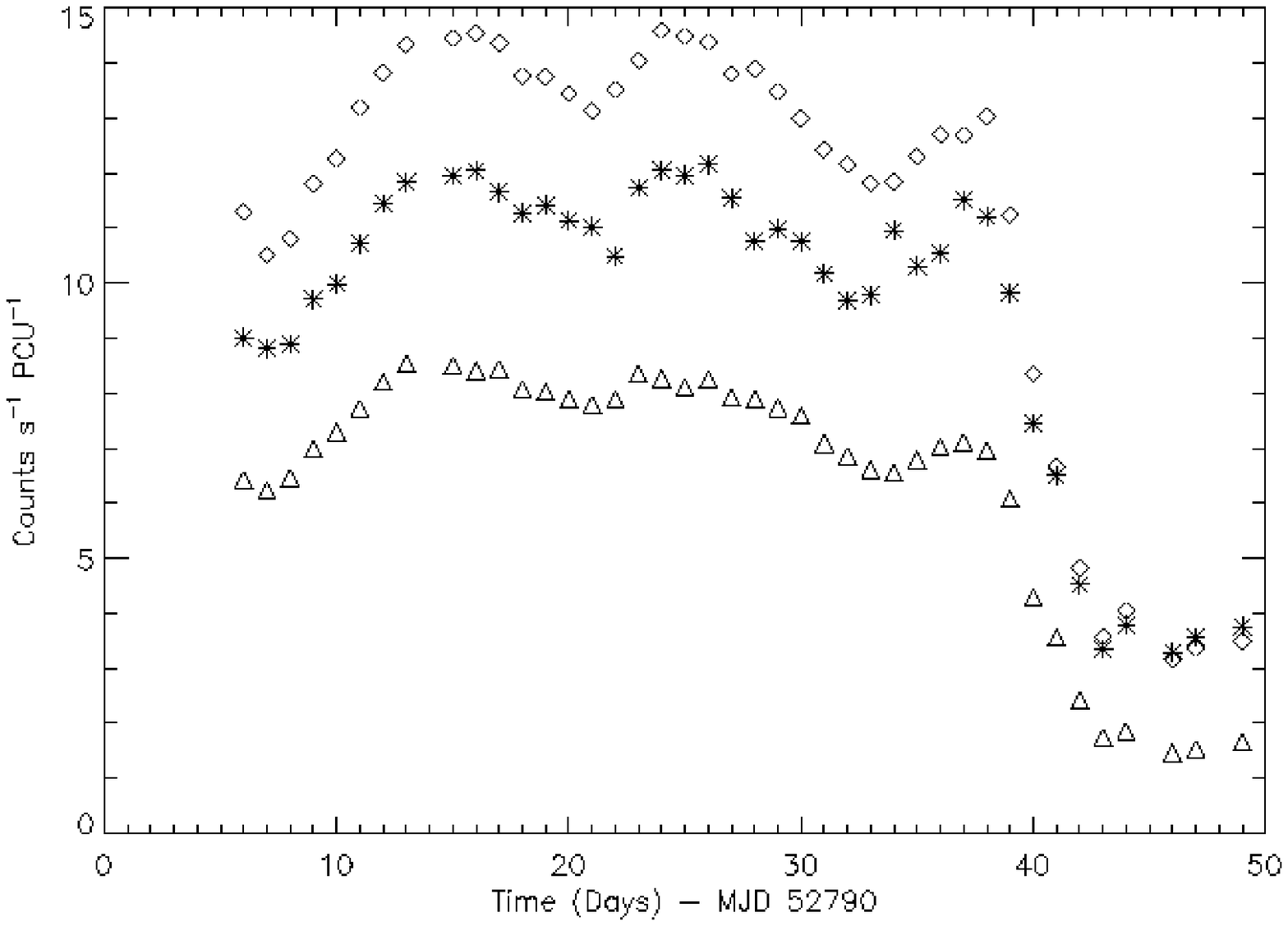}
\end{center}
\caption{Daily average countrate (excluding bursts) in different
energy bands, corrected 
for background.  {\it Stars}:  2-5 keV.  {\it Diamonds}:  5-10 keV.
{\it Triangles}:
10-20 keV.  }  
\label{f1}
\end{figure}

\begin{figure}
\begin{center}
\includegraphics[clip]{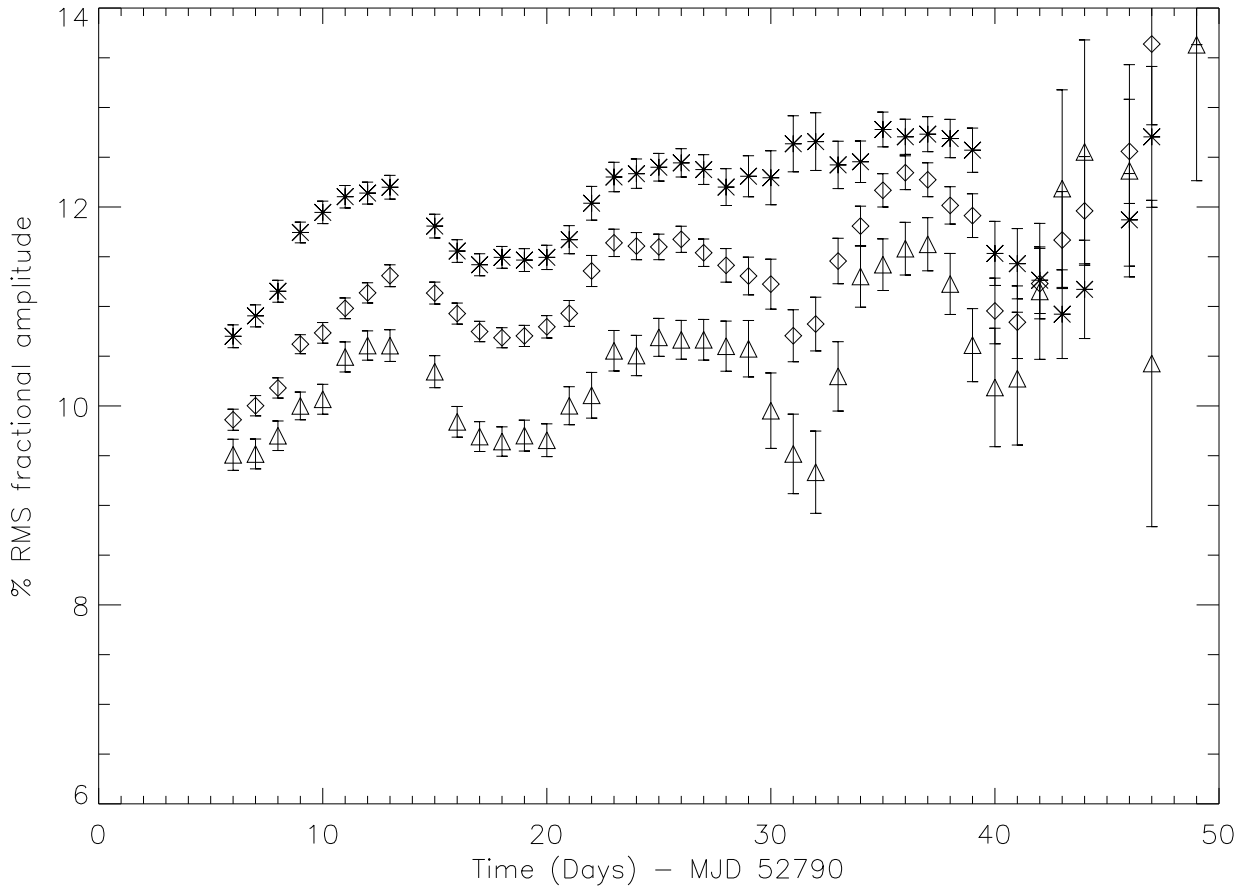}
\end{center}
\caption{Fractional amplitudes for the accretion-powered pulsations
for different energy bands, showing the variations over the outburst.  {\it Stars}:  2-5 keV.  {\it Diamonds}:
5-10 keV. {\it Triangles}: 10-20 keV. Each point is computed by
folding data for a 5 day 
period starting on the day indicated.  The drop in fractional
amplitude with energy persists throughout the outburst, although the
magnitude of the change does vary, particularly at the point where the
accretion rate starts to drop.}  
\label{f2a}
\end{figure}

\begin{figure}
\begin{center}
\includegraphics[clip]{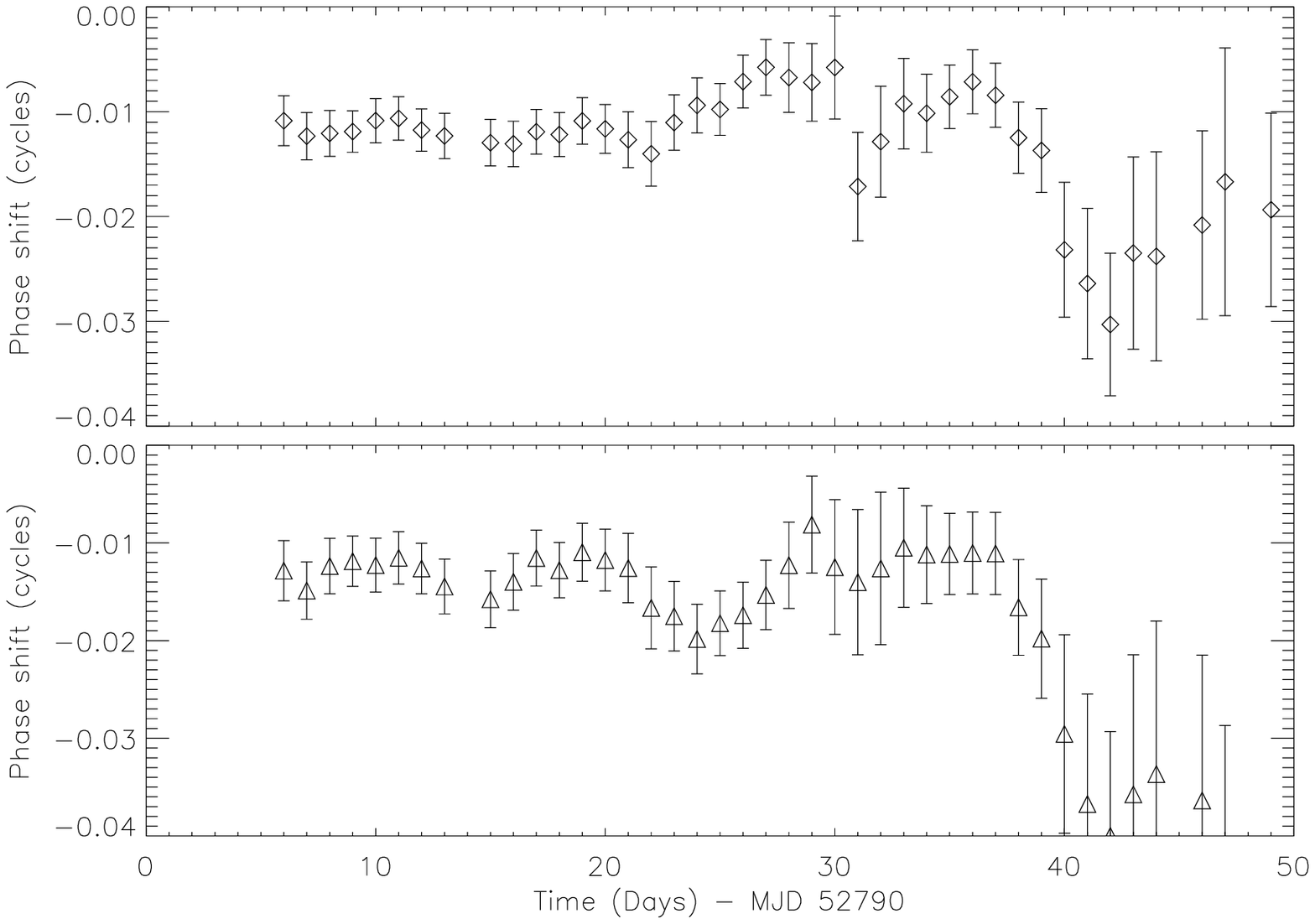}
\end{center}
\caption{Phase shifts compared to the phase for the 2-5 keV energy band for
the accretion-powered pulsations, showing the evolution over the outburst.{\it Top}: 5-10 keV. {\it
Bottom}: 10-20 keV. 
Negative values indicate that the pulse profile in the energy band being
studied leads the profile in the 2-5 keV band, with
0.01 cycles corresponding to 31.8$\mu$s. Each point is computed by
folding data for a 5 day 
period starting on the day indicated.}  
\label{f3}
\end{figure}

In order to correct for the accretion contribution recorded during the
bursts, we also need to know 
whether the properties of the persistent 
pulsations vary over the course of the outburst.  Figure \ref{f1}
shows the change in the countrate (excluding 
bursts) for the three energy bands
2-5 keV, 5-10 keV, and 10-20 keV.  The evolution of fractional amplitude (at the fundamental frequency)
is shown in Figure \ref{f2a}.  The trend of amplitude dropping with
energy is apparent over the whole outburst, although the magnitude of
the drop does vary between 1\% and 3\% RMS over the outburst.  The phase shifts between the different energy
bands are shown in Figure \ref{f3}. Again the soft lags persist 
throughout, but get noticeably larger at the end of the
outburst, when the accretion rate drops.  It would be interesting to
see if such a variation in phase lag with accretion rate is detected in
any of the other millisecond pulsars.  

\subsection{Burst oscillations}
\label{bur}

\begin{figure}
\begin{center}
\includegraphics[clip]{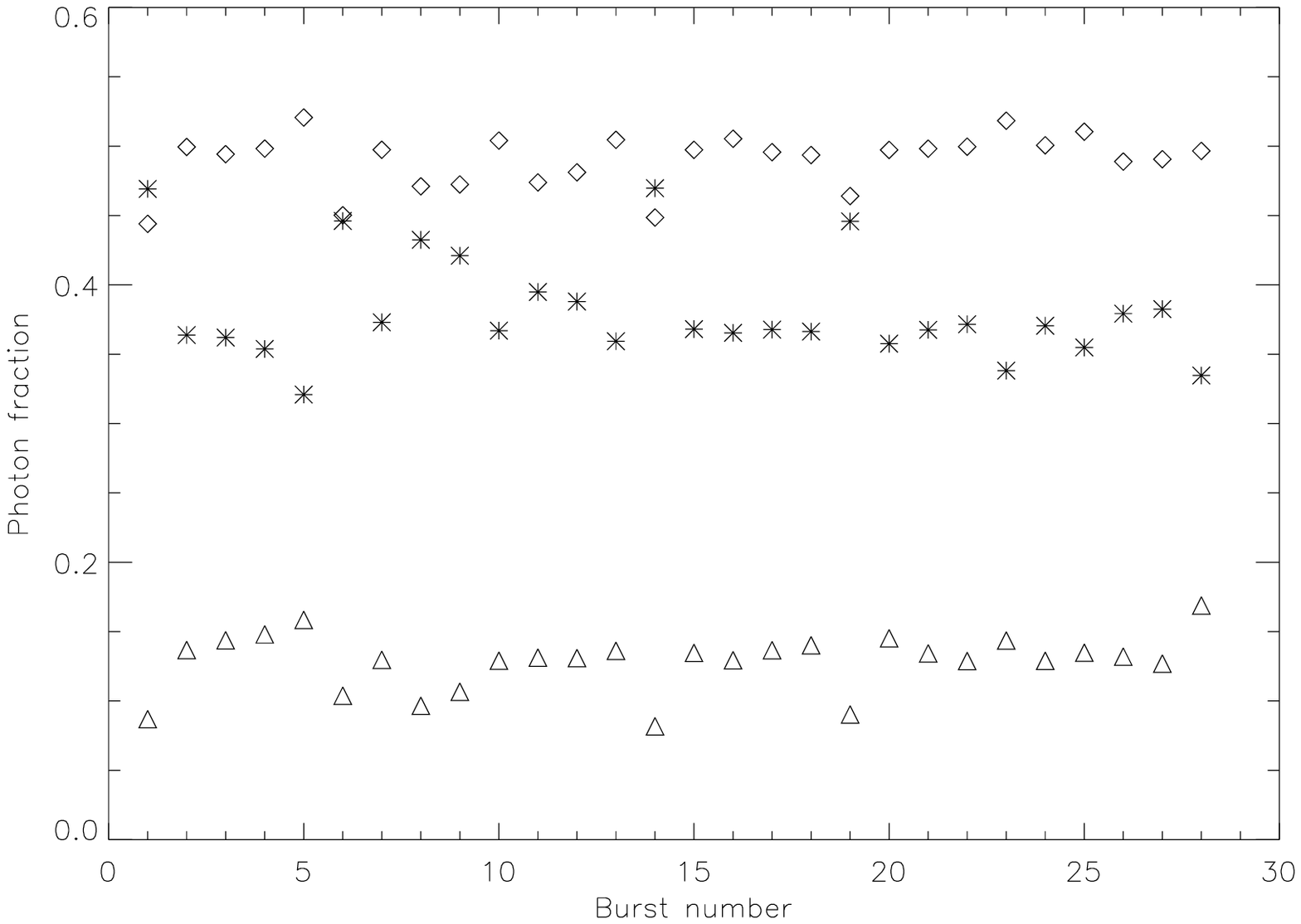}
\end{center}
\caption{Proportion of photons in the different energy bands during
each burst (corrected for background and an estimate of the
accretion-related emission). Proportions are roughly constant,
although for the faintest bursts  (1, 6, 8, 9, 14 and 19, see Figure
4 of Watts et al. 2005) the proportion of photons in the
highest energy bands drops. For the final, brightest burst, the
proportion in the highest energy band is higher. {\it Stars}:  2-5
keV. {\it Diamonds}:  5-10 keV. {\it Triangles}: 10-20 keV.}  
\label{f4}
\end{figure}

The proportion of photons in the different energy bands for each of
the bursts is shown in Figure \ref{f4}.  For most of the bursts
the proportions are very similar.  For the six faintest bursts, the
proportions in the higher energy bands drop, as might be expected given that these bursts have lower peak temperatures \citep{str03}.
Similarly, for the brightest burst, the proportion in the highest
energy band rises.  We define the duration of the burst as to be the time for which the countrate in a given energy band exceeds the ambient 
countrate by a factor of 1.5. Using this definition, burst duration is
 shorter at higher energies.  

\begin{figure}
\begin{center}
\includegraphics[clip]{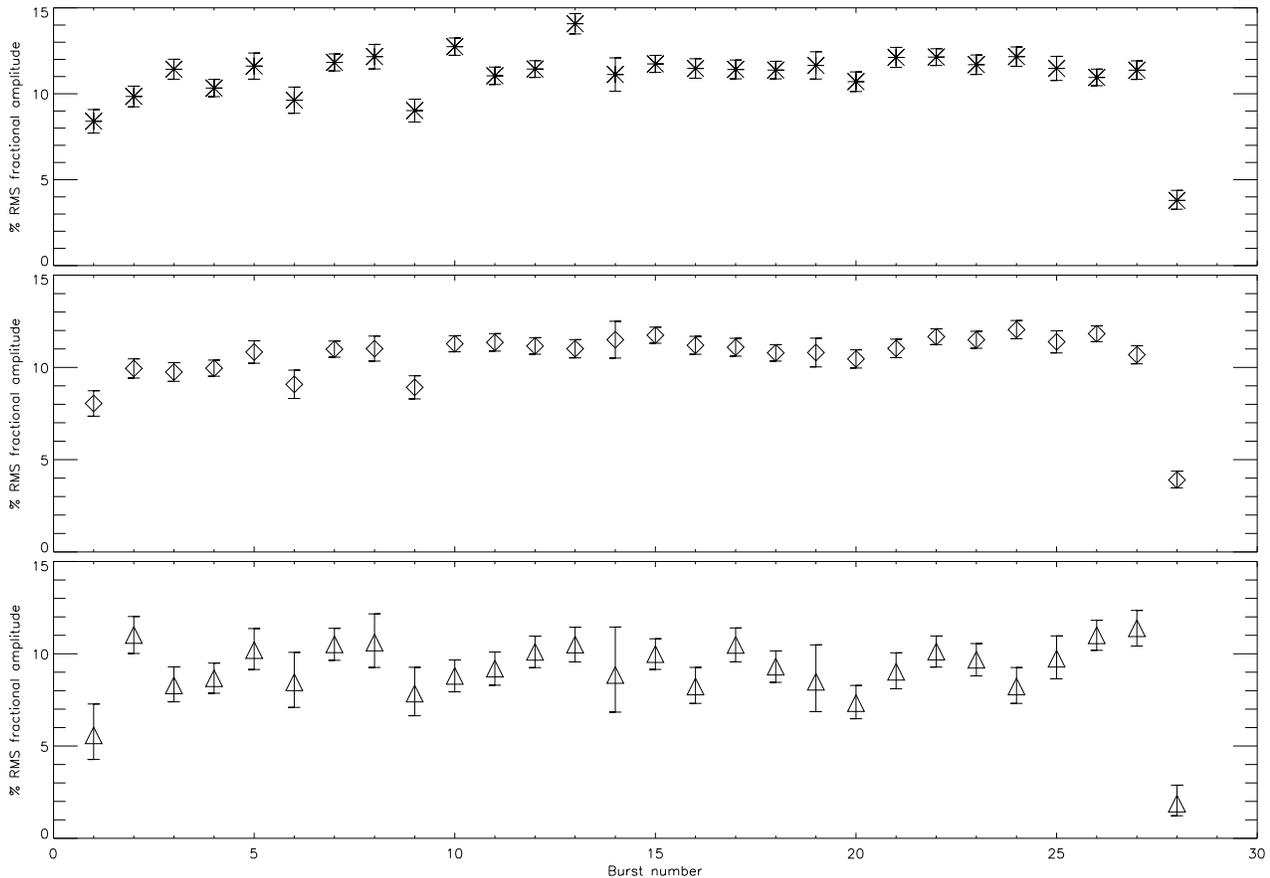}
\end{center}
\caption{Burst RMS fractional amplitude (\%) for different
energy bands, at the fundamental frequency of the pulsar. {\it Top}:  2-5
keV. {\it Middle}:  5-10 keV. {\it Bottom}: 10-20 keV. The scales on
each plot are the same to make clear the general drop in fractional amplitude with energy.}  
\label{f5}
\end{figure}

\begin{figure}
\begin{center}
\includegraphics[clip]{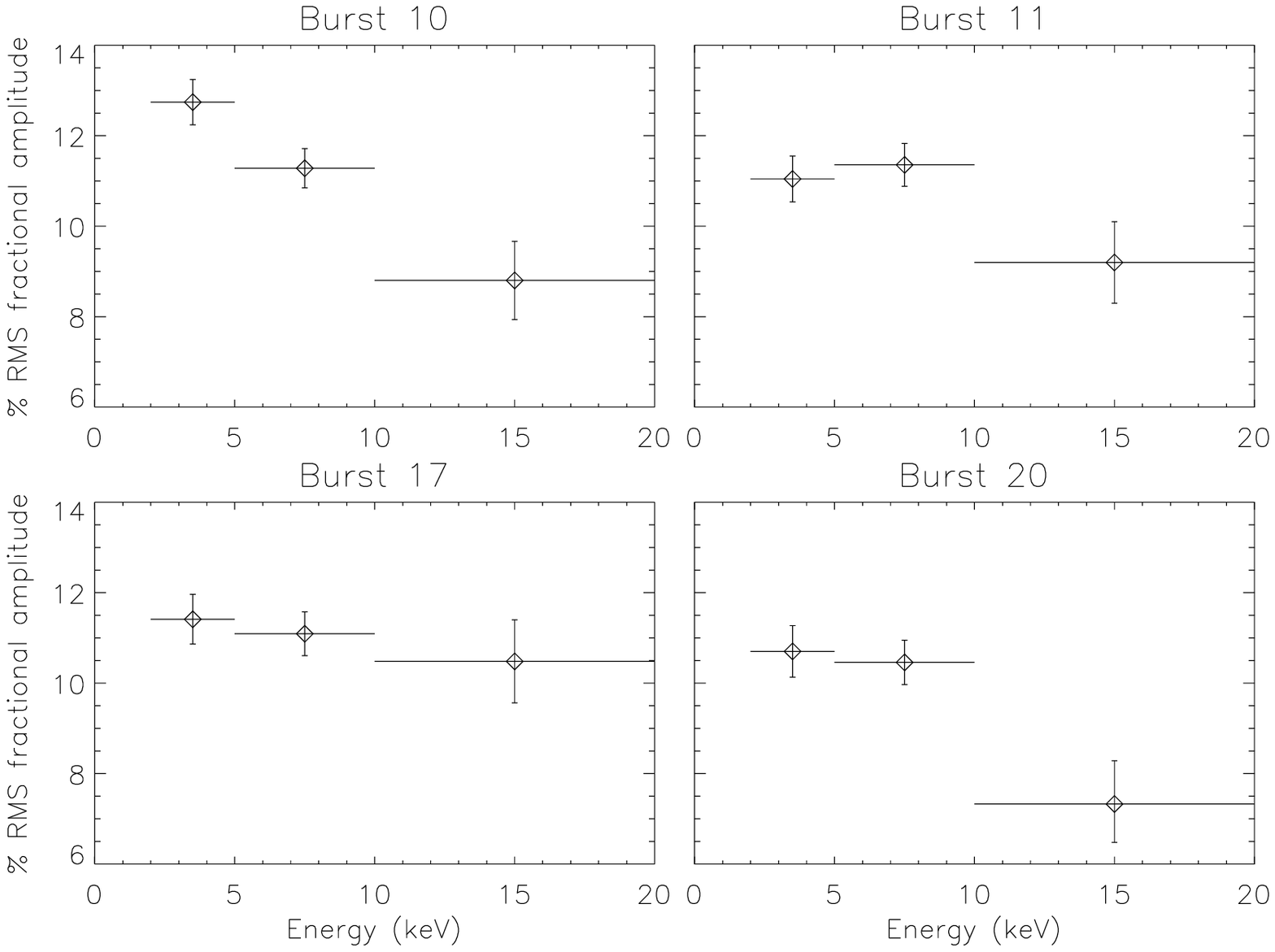}
\end{center}
\caption{Energy dependence of fractional amplitude for four example
bursts.  Burst numbers accord with those in Table \ref{bdata} of this paper and
Table 1 of \citet{wat05}. For bursts 10 and 17 a linear relation
between fractional amplitude and energy is a good fit, although the
gradients are very different. For bursts 11 and 20 the amplitude is
relatively constant between the 2-5 and 5-10 keV bands, dropping only
in the 10-20 keV band.}  
\label{f6}
\end{figure}

We start by considering each of the bursts separately, computing
fractional amplitudes and phase shifts for the three energy bands 2-5
keV, 5-10 keV and 10-20 keV.  The fractional amplitudes (at the
fundamental frequency) for all of the bursts are shown in Figure
\ref{f5}, and summarized in columns 2-4 of Table \ref{bdata}
(for more general data on each burst see Table 1 of \citet{wat05}). 
More detailed plots for four example bursts are shown in
Figure \ref{f6}.  We
find that burst amplitude is lower in the 10-20 keV band than in 
the 2-5 keV  band for 25 of the 28 bursts. Treating each burst as an
independent test, we find that we can rule out the hypothesis that
fractional amplitude is constant with energy at a level greater
than 3$\sigma$.  The hypothesis that fractional amplitude rises with
energy can be ruled out with an even higher degree of confidence.  

\begin{deluxetable}{cccccccccccc}
\tabletypesize{\scriptsize}
\tablecaption{{\bf Summary of burst properties}. Photon ratios (columns 5-7) and  accretion-corrected
  amplitudes (columns 8-10) are calculated assuming
 accretion rate and pulsation properties remain at pre-burst levels.  No error bars are given for columns 8-10 because of the uncertainties inherent in this assumption:  the error bars in columns 2-4 should be taken as a lower limit. \label{bdata}}
\tablewidth{0pt}
\tablehead{\colhead{Index} &  \multicolumn{3}{c}{RMS fractional amplitude (\%)} & \colhead{}  &
\multicolumn{3}{c}{$N_\mathrm{acc}$/$N_\mathrm{bur}$} & \colhead{}  &
\multicolumn{3}{c}{Accretion-corrected amplitudes} \\
\cline{2-4} \cline{6-8} \cline{10-12} \\ \colhead{}  &
  
  \colhead{2-5 keV} & 
  \colhead{5-10 keV} &\colhead{10-20 keV} & \colhead{} & \colhead{2-5 keV}  &
  \colhead{5-10 keV} &\colhead{10-20 keV} & \colhead{} & \colhead{2-5 keV}  &\colhead{5-10 keV}  &\colhead{10-20 keV}  
  }
  \startdata

  1 & 8.4 $\pm$ 0.7 & 8.0 $\pm$ 0.7 & $5.6^{+1.7}_{-1.3}$ && 0.24 &
  0.31 & 0.78 && 7.9 & 7.5  & 1.5 \\[4pt]

  2 & 9.8 $\pm$ 0.6 & 9.9 $\pm$ 0.5 & 11.0 $\pm$ 1.0 && 0.11 & 0.09 &
  0.15  && 9.8 & 10.0 & 11.3 \\[4pt]

  3 & 11.4 $\pm$ 0.6 & 9.8 $\pm$ 0.5 & $8.3^{+1.0}_{-0.9}$ && 0.15 &
  0.09 & 0.17  && 11.6 & 9.8  & 8.1 \\[4pt] 

  4 & 10.3 $\pm$ 0.5 & 10.0 $\pm$ 0.4 & 8.7 $\pm$ 0.8 && 0.19 & 0.11
  & 0.14  && 10.3 & 10.0  & 8.5 \\[4pt]

  5 & 11.6 $\pm$ 0.8 & 10.8 $\pm$ 0.6 & $10.2^{+1.2}_{-1.0}$ && 0.17
  & 0.12 & 0.19  && 11.7 & 10.9 & 10.4 \\[4pt]

  6 & 9.6 $\pm$ 0.8 & 9.1 $\pm$ 0.8 & $8.5^{+1.6}_{-1.4}$ && 0.35 &
  0.34 & 0.89 && 8.9 & 8.5 & 6.9 \\ [4pt]

  7 & 11.8 $\pm$ 0.5 & 11.0 $\pm$ 0.4 & 10.5 $\pm$ 0.9 && 0.20 & 0.14
  & 0.20 && 11.9 & 11.1 & 10.6 \\ [4pt]

  8 & 12.2 $\pm$ 0.7 & 11.0 $\pm$ 0.7 & $10.6^{+1.6}_{-1.4}$  && 0.31
  & 0.34 & 0.67 && 12.3 & 11.2 & 10.8 \\ [4pt]

  9 & 9.0 $\pm$ 0.7 & 8.9 $\pm$ 0.6 & $7.8^{+1.4}_{-1.2}$ && 0.21 &
  0.23  & 0.41 && 8.3 & 8.4 & 6.7 \\ [4pt]

  10 & 12.7 $\pm$ 0.5 & 11.3 $\pm$ 0.4 & 8.8 $\pm$ 0.9 && 0.19 & 0.16
  & 0.22 && 12.8 & 11.3 & 8.4 \\ [4pt]

  11 & 11.0 $\pm$ 0.5 & 11.4 $\pm$ 0.5 & 9.2 $\pm$ 0.9 && 0.18 & 0.12
  & 0.22 && 10.8 & 11.4 & 8.9 \\ [4pt]
 
  12 & 11.4 $\pm$ 0.5  & 11.2 $\pm$ 0.4 & 10.1 $\pm$ 0.9 && 0.19 &
  0.14 & 0.26 && 11.3 & 11.1 & 9.9 \\[4pt] 

  13 & 14.1 $\pm$ 0.6 & 11.0 $\pm$ 0.5 & 10.5 $\pm$ 0.9 && 0.14 &
  0.17 & 0.33 && 14.4  & 10.9 & 10.4 \\[4pt]

  14 & 11.1 $\pm$ 1.0 & 11.5 $\pm$ 1.0 & $8.9^{+2.6}_{-2.0}$ && 0.26 &
  0.26 &0.51 && 10.9 & 11.5 & 7.8 \\[4pt]

  15 & 11.7 $\pm$ 0.5 & 11.7 $\pm$ 0.4 & 10.0 $\pm$ 0.8 && 0.17 &
  0.13 & 0.23  && 11.8 & 11.8 & 10.1 \\ [4pt]

  16 & 11.5 $\pm$ 0.6 & 11.2 $\pm$ 0.5 & $8.2^{+1.0}_{-0.9}$ && 0.16
  & 0.14 & 0.18 && 11.4 & 11.2  & 7.9 \\[4pt]

  17 & 11.4 $\pm$ 0.5 & 11.1 $\pm$ 0.5 & 10.5 $\pm$ 0.9 && 0.20 &
  0.13 & 0.24 && 11.4 & 11.1 & 10.6 \\[4pt]

  18 & 11.4 $\pm$ 0.5 & 10.8 $\pm$ 0.4 & 9.3 $\pm$ 0.8 && 0.13 & 0.11
  & 0.14 && 11.4  & 10.8 & 9.3 \\[4pt]

  19 & 11.7 $\pm$ 0.8 & 10.8 $\pm$ 0.8 & $8.5^{+2.0}_{-1.6}$ && 0.41
  & 0.44 & 0.58 && 11.9 & 10.9  & 8.1 \\[4pt]

  20 & 10.7 $\pm$ 0.6 & 10.5 $\pm$ 0.5 & $7.3^{+1.0}_{-0.8}$ && 0.17
  & 0.12 & 0.15 && 10.6 &10.4 & 6.9 \\[4pt]

  21 & 12.1 $\pm$ 0.6 & 11.0 $\pm$ 0.5 & $9.0^{+1.0}_{-0.9}$ && 0.13
  & 0.12 & 0.22 && 12.1 & 11.1 & 9.0 \\[4pt]

  22 & 12.1 $\pm$ 0.5 & 11.7 $\pm$ 0.4 & 10.1 $\pm$ 0.8 && 0.20 &
  0.16 & 0.27 && 12.2  & 11.7 & 10.0 \\[4pt]

  23 & 11.7 $\pm$ 0.6 & 11.5 $\pm$ 0.5 & 9.7 $\pm$ 0.9 && 0.11 & 0.09
  & 0.18 && 11.6 & 11.5 & 9.5 \\[4pt]

  24 & 12.2 $\pm$ 0.6 & 12.0 $\pm$ 0.5 & $8.2^{+1.0}_{-0.9}$ && 0.16
  & 0.13 & 0.23 && 12.1  & 12.1 & 7.6 \\[4pt]

  25 & 11.5 $\pm$ 0.7 & 11.4 $\pm$ 0.6 & $9.7^{+1.2}_{-1.1}$ && 0.21
  & 0.14 & 0.28 && 11.4 & 11.4 & 9.6 \\[4pt]

  26 & 10.9 $\pm$ 0.5 & 11.8 $\pm$ 0.4 & 11.0 $\pm$ 0.8 && 0.14 &
  0.11 & 0.30 && 10.8 & 11.8 & 10.8 \\[4pt]

  27 & 11.4 $\pm$ 0.5 & 10.7 $\pm$ 0.5 & 11.4 $\pm$ 1.0  && 0.19 &
  0.10 & 0.23 && 11.1  & 10.5 & 11.0 \\[4pt]

  28 & $3.8^{+0.6}_{-0.5}$ & $3.9^{+0.5}_{-0.4}$ &
  $1.9^{+1.0}_{-0.7}$ && 0.04  & 0.02 & 0.02 && 3.5 & 3.7 & 1.8
  \\
  \enddata
  \end{deluxetable}

\begin{figure}
\begin{center}
\includegraphics[clip]{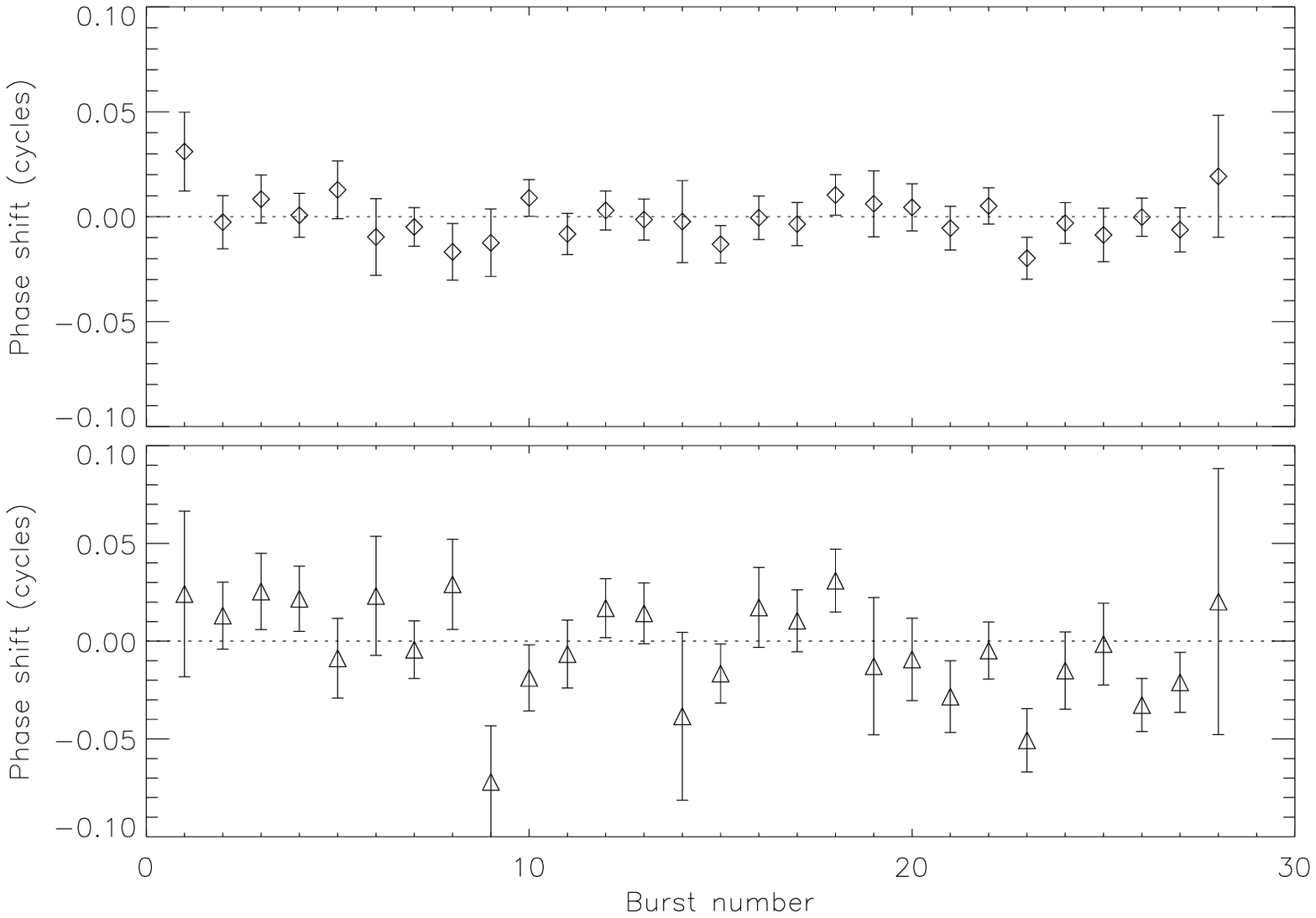}
\end{center}
\caption{Phase shifts compared to the phase for the 2-5 keV band for
the bursts.{\it Top}: 5-10 keV. {\it
Bottom}: 10-20 keV. 
Negative values indicate that the pulse profile in the band being
studied leads the profile in the 2-5 keV band, with
0.01 cycles corresponding to 31.8$\mu$s. }  
\label{f7}
\end{figure}

The phase shifts between the different energy bands for each burst are
shown in Figure
\ref{f7}.   There is no evidence for
either hard or soft lags;
the data are compatible with there being no phase shift.  

\begin{figure}
\begin{center}
\includegraphics[clip]{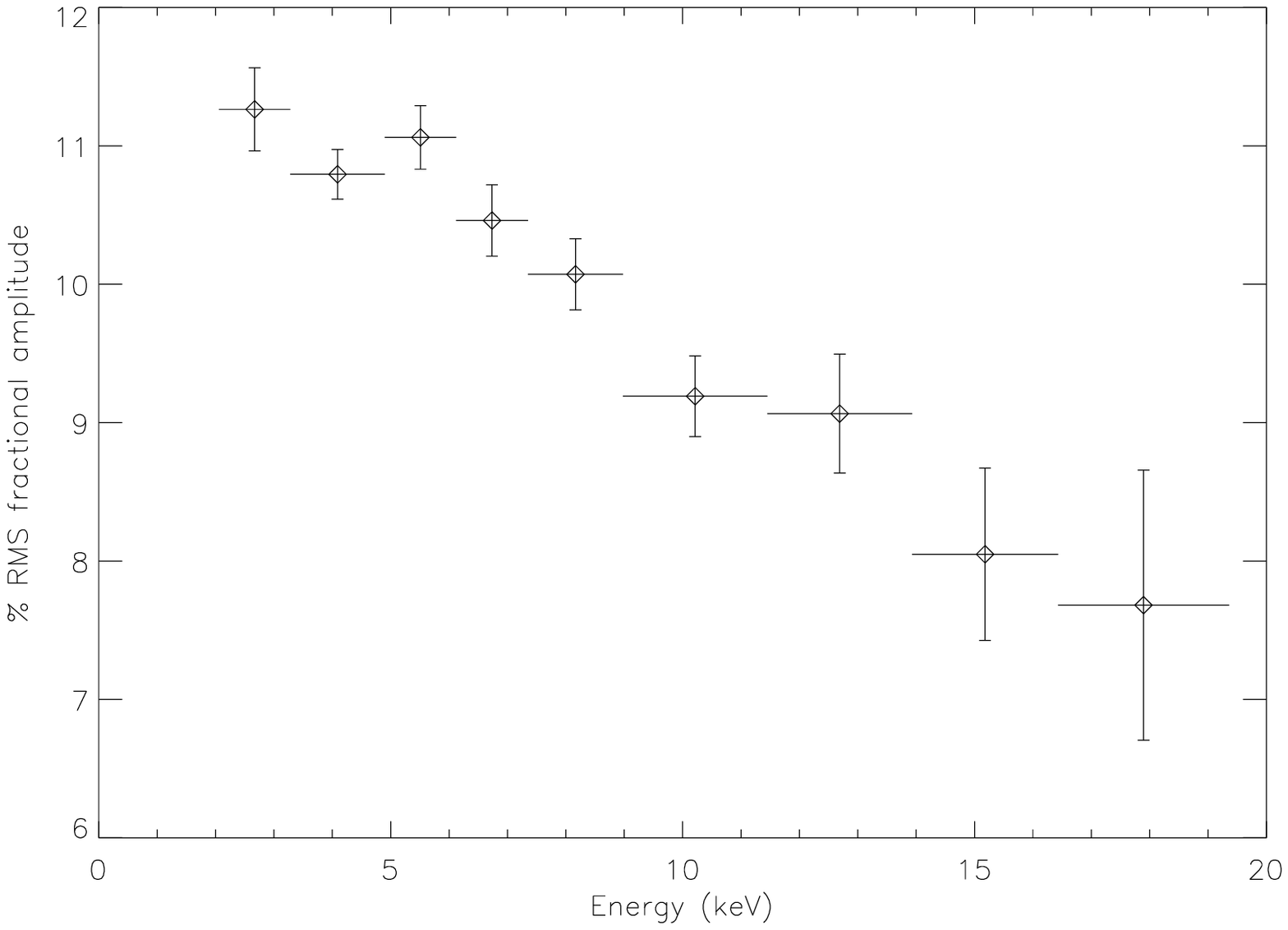}
\end{center}
\caption{The dependence of fractional amplitude on energy for the
bursts, computed from a folded profile of data from all of the
bursts.  No attempt has been made to correct for the accretion
contribution (see discussion in the text). }  
\label{f8a}
\end{figure}

\begin{figure}
\begin{center}
\includegraphics[clip]{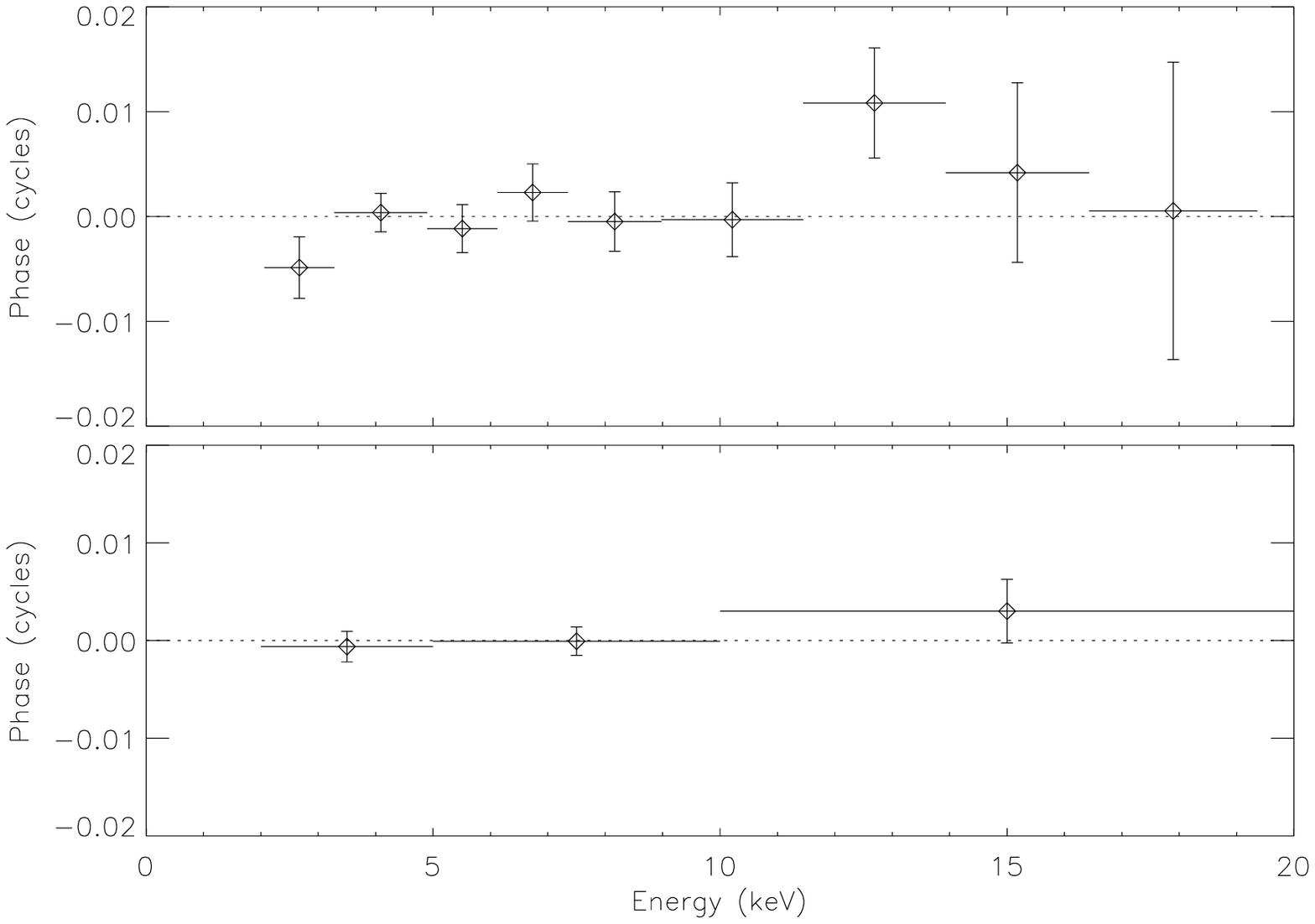}
\end{center}
\caption{Phases calculated for the folded set of bursts, relative to
the best fitting constant phase.  The top panel shows a higher energy
resolution than the lower panel. Both sets of data are consistent with
there being no  phase shift between the different energy bands.  
Note that 0.01 cycles corresponds to 31.8$\mu$s. }  
\label{f8b}
\end{figure}

In order to verify these results,
we folded together data from all of the bursts to generate a
combined profile.  We find a drop of fractional
amplitude with energy of $\approx$  3\% RMS over
the 2-20 keV band, a steeper drop than that seen in the persistent
pulsations (Figure \ref{f8a}).  The phase shifts of the combined
profile, for two energy binnings, are shown in Figure \ref{f8b}.  The data are
 consistent with there being no phase shift, and we can rule
out at the 3$\sigma$ level hard or soft lags of greater than 0.015
cycles over the 2-20 keV band.  We would have been sensitive to lags
of the magnitude seen in the accretion-powered pulsations were such lags
present.  

\section{Discussion}
\label{disc}

The burst oscillations of J1814 exhibit a  drop in fractional
amplitude with energy of a few percent RMS across the 2-20 keV band.
In addition the bursts show no evidence for phase shifts between soft
and hard photons;  we can rule out at the 3$\sigma$ level shifts greater
than 0.015 cycles over the 2-20 keV band.  In this section we discuss these
results in the context of the various burst oscillation models.  

\subsection{Consistency with existing burst oscillation models}

We start by considering models where surface modes cause the
brightness asymmetry.  The mode models discussed by \citet{hey05,
lee05} and \citet{pir06} 
all predict a rise of amplitude with energy,  
although the gradient of the rise depends on various neutron star
parameters.  Since we can rule out a rise, or even a constant
amplitude, with a high degree of confidence, the amplitude data
are inconsistent with these models.  The phase lag data do not provide
any additional constraint, since the models of (for example)
\citet{lee05} predict anything from hard lags of 0.04 cycles to soft
lags of 0.1 cycle depending on the neutron star properties.  Although we can rule
out lags greater than 0.015 cycles there are still models that would
fit the phase lag data.

The other class of models that have been studied in any detail are hot spot models.  These models are purely phenomenological, in that the physics that might give rise to a hot spot remains as yet unclear.   Like the mode models, the simple one-temperature 
hot spot models studied by \citet{mun03} also predict a rise of  
amplitude with energy, although again
the magnitude of the rise 
depends on the neutron star properties.  The models studied by Muno
et al. also predict soft lags of least 0.02 cycles over the RXTE
energy band.  Any phase lags present in the burst oscillations of J1814
are smaller than this limit.  

We are therefore left with several options.  One possibility
is that the burst oscillation properties are consistent with existing
mode/hot spot models but are masked by the presence of
accretion-powered pulsations.  We examine this in more detail in Section \ref{mask}, and show that the accretion process
would have to be disrupted substantially by the bursts to explain the
observations. A second option is that existing models 
are not adequate to explain the observations.  In Section \ref{tgrad} we explore whether a more complex surface temperature gradient might lead to consistency with the
observations.

\subsection{Masking by the accretion pulsations}
\label{mask}
Unless the accretion process is completely disrupted by the burst, the
measured fractional amplitude will contain contributions from both
accretion and burst pulsations.  Separating the two contributions is
not trivial because in this source they are phase-locked
\citep{str03}.   The persistent pulsations of J1814 do indeed show a
drop in fractional 
amplitude with energy of 1-3\% RMS over the 2-20 keV energy band
(during the main portion of the outburst), so masking is not
inconceivable.  To proceed we need to estimate the magnitude of the accretion
contribution to Equation (\ref{prop}).

We start by making the simple assumption that both the accretion rate
and the amplitude/energy relation of the accretion-powered pulsations
remain in their pre-burst state, as presented in Section \ref{acc},
during a burst.   Columns 5-7 of Table 
\ref{bdata} give an estimate of the ratio 
of accretion to burst photons over the duration of the burst, computed from
the daily average accretion rate at the time of the burst.  In Columns
8-10 we combine this estimate with the daily average fractional
amplitude for the accretion-powered pulsations to estimate
$r_\mathrm{bur}$, using Equation (\ref{prop}).  Although the
accretion correction does make some small quantitative changes, it does not
change the
qualitative behaviour with energy for any of the bursts.  Unless the
accretion contribution differs dramatically at the time of the bursts
from that observed normally, the burst fractional amplitude still falls
with energy.  So let us now consider the magnitude of changes to the
accretion pulsations, compared 
to the properties measured immediately prior to each bursts, that
would be necessary to mask burst pulsations whose
fractional amplitude rises with energy. 

If the accretion rate remains
at the pre-burst level,  the fractional 
amplitude of the accretion  
pulsations must fall more steeply with energy during the burst than it
does normally.  Either the amplitude at low energies must rise, or the
amplitude at higher energies fall, or both.   For most of the bursts
the ratio of accretion to burst photons is less than 0.3 (Table
\ref{bdata}).  The fractional amplitude would have to change by
several \% RMS to have the desired effect.  Whether the burst process
could lead to this level of disruption is not known, but it seems
unlikely.   A smaller change
in fractional amplitude would be required if the 
accretion rate rose during a burst, boosting the ratio of
accretion to burst photons.  The effect of bursts on accretion rate
is however unclear.  Naively one might expect the radiation pressure to hinder the
accretion process.   But the radiation may also remove angular
momentum, increasing the accretion rate \citep{wal89, wal92, mil93,
mil95, bal05}.  A substantial rise in accretion rate seems however unlikely,
since the burst spectra are well-fit by a single temperature
blackbody, and there is no spectral evidence for an enhanced accretion
component.  

The one-temperature hot spot models, and many of the mode models,
predict burst oscillations with detectable soft lags.  This
could only be masked by the accretion process if the accretion-powered pulsations
exhibit
hard lags during bursts.  This would be a reversal of the normal
behaviour, requiring substantial disruption of the accretion process by
the burst.  Whether or not this is plausible is a topic for future
study.  

\subsection{Hot spot with temperature gradient}
\label{tgrad}

The evidence presented in \citet{str03} and \citet{wat05} suggests
that the most promising model for the burst oscillations of J1814 is a
hot spot that develops at the magnetic footpoint, perhaps due to
a disparity in fuel build-up in this region.  We will therefore focus
on the hot spot model in more detail, and ask whether some simple
alterations to the one-temperature hot spot models of \citet{mun03}
might enable us to explain the energy-dependences reported in this
paper.  

Although most modelling to date has focused on uniform, single
temperature hot spots, several processes involved in burst generation
would seem to favour some degree of temperature variation across a hot
spot. For example, we expect that burning must spread in some fashion
such that regions which ignite first will have a longer time to cool
than portions of a hot spot which ignite later.  The time asymmetry
introduced by spreading can thus lead to a temperature asymmetry.
Alternatively, if accretion is substantially mediated by the magnetic
field, then the quantity of fuel could vary in a systematic fashion in
the vicinity of the magnetic footpoint.  Such a fuel asymmetry could
also impart a temperature gradient within the hot spot.

In order to explore the impact of a temperature gradient we have carried out model calculations using circular hot spots 
with a temperature profile across the spot.  It is not our intention to
fit detailed models to the data, only to examine whether a temperature
gradient might allow us to fit the amplitude-energy relation.   We use the model for emission of 
photons from a rotating neutron star surface described by
\citet{str04}. The model includes bending of photon paths in a
Schwarzschild  
geometry as well as relativistic beaming and gravitational red-shifts.
Each surface element within the hot spot is assumed to emit, locally,
a black body spectrum. The temperature distribution is azimuthally
symmetric around the center of the hot spot, and is assumed to vary
with the angle, $\alpha$, measured from the central axis of the spot.
For the purposes of this exploratory calculation we use a simple
linear variation of the temperature with angle, $kT = kT_0 (1 +
\Delta\alpha / \alpha_0 )$, where $kT_0$, $\Delta$, $\alpha$ and $\alpha_0$ are
the central temperature, fractional temperature change, half-angle of
the emitting point (measured from the spot centre) and maximum
half-angular 
size of the hot spot, respectively.    We
compute a model photon energy spectrum at infinity,  
and then apply an RXTE/PCA response matrix to obtain count rates
versus PCA energy channel.  We then compute amplitudes as a function
of energy channel that can be compared with the real data. 
 
\begin{figure}
\begin{center}
\includegraphics[clip]{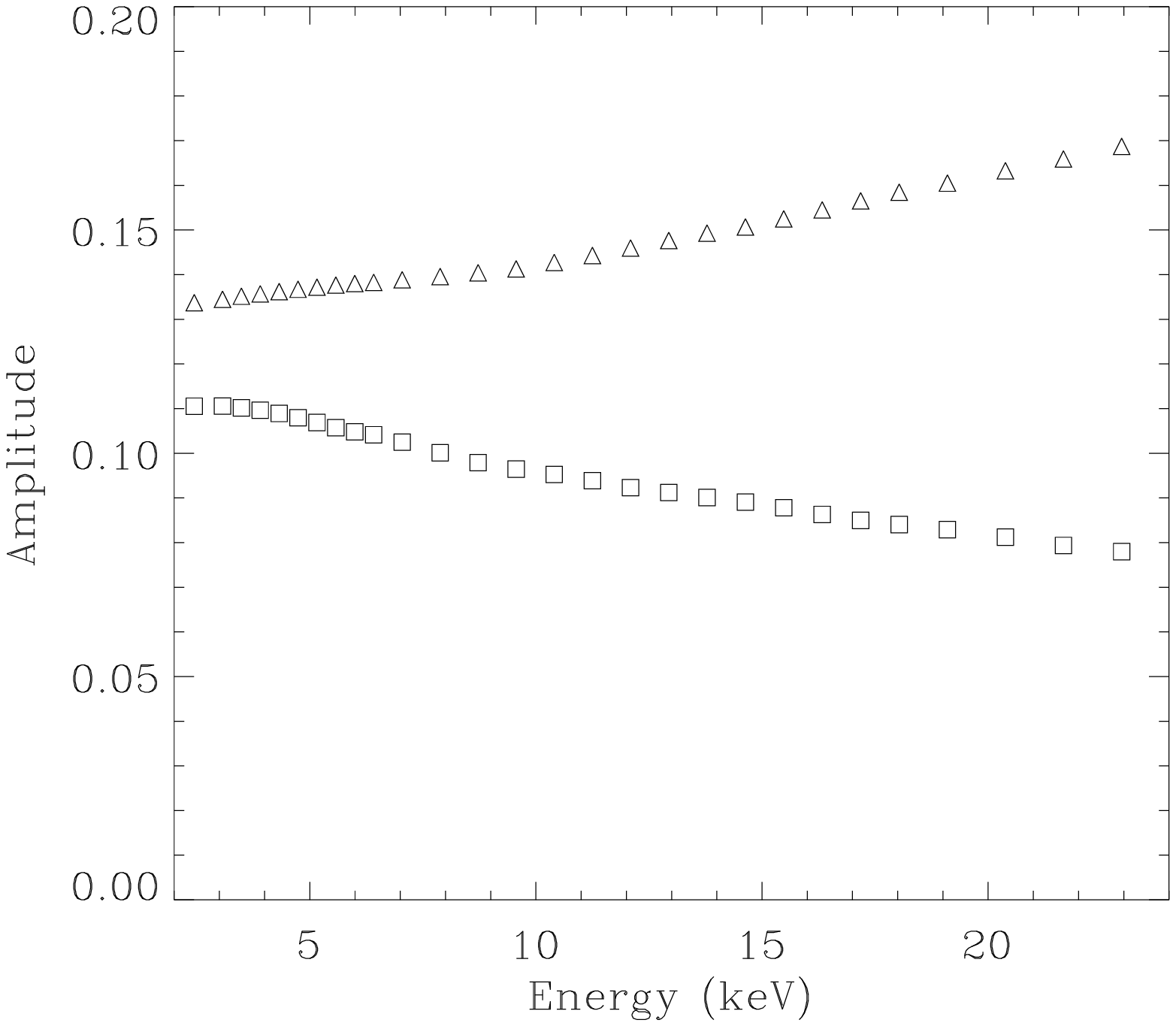}
\end{center}
\caption{Rotational modulation amplitude as a function of energy for two circular
hot spots; one with uniform temperature, and one with a linear variation in temperature across the spot.  In
these calculations we used a spot size (half angle) of $138^\circ$, a
location of the spot center of $80^\circ$, and an observer viewing
angle of $45^\circ$.  The temperature at the center of the spot is 2.9
keV, whilst the temperature of the rest of the star is 0.8 keV
(although there is little change if this is lower).  The upper trace
shows the result for a uniform temperature hot spot.  The lower trace
shows the result for $\Delta = 0.1$, so that the outer boundary of the
spot is at 3.2 keV.}
\label{f9}
\end{figure}

Two example models are shown in Figure \ref{f9}.  We
have selected a geometry that gives an amplitude close to the observed
values. As expected, and as shown by other authors, a uniform spot
emitting a 
typical burst spectrum produces an increasing amplitude with energy in
the PCA bandpass. Since in general a smaller hot spot produces a
larger modulation amplitude, one would expect that in order to produce
a drop in amplitude with increasing energy the temperature should
increase {\it away} from the center of the spot. That is, the edge
should be hotter than the center. This expectation is borne out in our
modeling.  In the example shown, adding a fractional temperature
change of $\approx$ 10\% produces a drop in amplitude with energy of similar
magnitude to that observed.

We do not claim that the above geometry is unique in fitting this
aspect of the data, and it is not clear that it would survive other
constraints such as pulse profile fitting (see for example
\citet{bha05}).  However, it does
demonstrate that temperature 
gradients merit further study.  How such systematic variations in
temperature might arise is not clear, although we note that an
asymmetry associated with spreading from the center of a hot spot
would tend to produce a hotter edge (as the center has longer to
cool).  Constraints from theory on the magnitude and distribution of likely temperature variations would be extremely helpful.

\section{Conclusions}

In this paper we have presented an analysis of the energy dependence
of burst oscillations from the accreting millisecond pulsar J1814.
Our results are intriguing.  The fractional amplitude of the
pulsations falls with energy across the 2-20 keV RXTE band; we are
able to rule out the amplitude being constant, or rising with energy,
at a level greater than 3$\sigma$.  This mirrors the behaviour seen in
the persistent accretion-powered pulsations, but differs from that
seen for the burst oscillations of the non-pulsing LMXBs by
\citet{mun03}.  The persistent pulsations, like those of the other
accreting millisecond pulsars, show soft lags of 0.015 cycles over the
2-20 keV band; for the burst oscillations we can rule out lags or
leads of this magnitude at the 3$\sigma$ level. 

The energy dependence of J1814's burst oscillations differs from that
reported for LMXB burst oscillations in the most comprehensive study
of these objects to date \citep{mun03}.  It joins a list of properties
that seem to differ between the two classes of objects.  For the
LMXBs, burst oscillations are only seen at high accretion rates; for
the pulsars burst oscillations are seen even though the accretion rate
never gets high \citep{mun04}.  LMXB burst oscillations show frequency
shifts of up to 5Hz (see for example \citet{mun01}); for the pulsars
frequency variation is only seen in the rises of those bursts that
show PRE \citep{cha03, wat05, bha06}.  The LMXB burst oscillations also show
amplitude variations \citep{mun02}; for J1814 no such variations are
seen except during PRE \citep{wat05}.  Finally, the pulsar burst
oscillations have detectable harmonic \citep{cha03, str03},
whereas the LMXB oscillations do not (\citet{mun02}, although see \citet{bha05}, which presents evidence for harmonic content in the rising phase of bursts from 4U 1636-536).   

The differing properties suggest that the mechanism responsible for
the burst oscillations may differ between the two classes of objects.
Further study, however, is clearly required.   There is now a much
larger sample of LMXB burst oscillations in the RXTE archive, and it
would be advisable to check whether the trends found in the studies by
Muno and collaborators still hold for a larger sample.   The energy
dependence of the burst oscillations of the accreting millisecond
pulsar J1808 also remains to be analysed.   Our study also highlights
the need to look at variations in properties over the course of the
outburst; both accretion-powered and burst oscillation properties were
found to vary for J1814.  The technique of folding together multiple
bursts, for example (used both in this paper and in the previous study
by \citet{mun03}), could obscure interesting variations.   

The drop in fractional amplitude with energy observed in the burst
oscillations of J1814 is inconsistent with the predictions of both
surface mode and one-temperature hot spot models.  If these models are
correct then there must be substantial changes in the accretion
process, triggered by the burst, that cause the accretion pulsations
to mask this property of the thermonuclear pulsations.  If we assume
that the accretion process does not alter significantly during the
burst then we must seek an alternative theory.  One possibility is
that the hot spot possesses a temperature gradient.   In Section
\ref{tgrad} we examined the magnitude of the gradient that might be
required to explain the observations.  It is substantial:  a rise of
$\sim$ 10\% from center to edge.  How such a gradient might be
generated and whether it can meet other constraints posed by, for
example, the pulse profile, are both interesting questions for future work.
Other processes such as beaming due to a corona, however, may have to be invoked to explain the observations.

\section{Acknowledgments}

We are grateful to Tony Piro for encouraging us to look at this issue,
and to Craig Markwardt for providing the orbital ephemeris for XTE
J1814-338.  We would also like to thank the referee, Mike Muno, for
helpful comments and for discussing his own unpublished analysis of
this data.  ALW acknowledges partial support from the European Union
FP5  
Research Training Network 'Gamma-Ray Bursts:  An Enigma and a
Tool'. This research has made use of data obtained from the High 
Energy Astrophysics Science Archive Research Center (HEASARC)
provided by NASA's Goddard Space Flight Center.

\end{document}